\documentclass[sigplan,10pt,nonacm]{acmart}
\renewcommand\footnotetextcopyrightpermission[1]{}
\usepackage{booktabs}
\usepackage{tikz}
\usetikzlibrary{arrows.meta,positioning,calc,fit,shapes.misc}
\usepackage{amsmath}
\usepackage{filecontents}
\usepackage[normalem]{ulem}
\usepackage{graphicx}
\usepackage{pifont}
\usepackage{balance}
\usepackage{multirow}
\usepackage{enumitem}
\usepackage{xspace}
\usepackage[]{hyperref}
\usepackage{algorithm}
\usepackage{algpseudocode}
\usepackage{listings}
\usepackage{xcolor}
\usepackage{makecell}
\usepackage{subcaption}
\usepackage{array}
\newcommand{\origunderscore}{}
\let\origunderscore\_
\DeclareRobustCommand{\code}[1]{{\texttt{\renewcommand{\_}{\origunderscore\hspace{0pt}}#1}}}

\usepackage{titlesec}
\titlespacing*{\section}{0pt}{*0.9}{*0.9}
\titlespacing*{\subsection}{0pt}{*0.9}{*0.9}
\titlespacing*{\subsubsection}{0pt}{*0.9}{*0.9}

\newcommand{\sys}{VUDA\xspace}

\newif\iffinalcheck
\finalchecktrue

\newcommand{\myparagraph}[1]{\vspace{3pt}\noindent\textbf{\emph{#1}}}

\newcommand*\circled[1]{\tikz[baseline=(char.base)]{
            \node[shape=circle,draw,inner sep=1pt,color=white,fill=black] (char) {#1};}}
\newcommand*\whitecircled[1]{\tikz[baseline=(char.base)]{
            \node[shape=circle,draw,inner sep=1pt,color=black,fill=none] (char) {#1};}}

  {\begin{list}{\labelitemi}{\itemsep1pt \topsep2pt \parsep0.00in
  \partopsep=0pt \leftmargin1em}}%
  {\end{list}}

\newcommand{\yes}{\ding{51}}
\newcommand{\no}{\ding{55}}
\newcommand{\partialmark}{$\triangle$}

\definecolor{deepblue}{RGB}{0, 0, 139}
\definecolor{gfxcolor}{RGB}{66,133,244}
\definecolor{compcolor}{RGB}{219,68,55}

\begin{document}
\date{}

\title[VUDA]{VUDA: Enabling Controlled Spatial Sharing of Graphics and Compute on NVIDIA GPUs}

\author{Bin Xu, Pengfei Hu, Jinyu Gu*, and Haibo Chen}
\affiliation{%
  \institution{Shanghai Jiao Tong University}
  \city{Shanghai}
  \country{China}
}
\renewcommand{\shortauthors}{Xu et al.}

\begin{abstract}
Graphics and compute increasingly share a single GPU in embodied AI
simulators, AI-enabled games, and VR systems.
Running these workloads concurrently can improve utilization, but
contention can also compromise rendering and inference latency.
Effective sharing therefore requires control over both concurrency and
resource allocation.
Native CUDA and Vulkan runtimes complicate this task: they place work in
separate scheduling domains, while compute-oriented resource controls do
not govern the graphics pipeline.

We present \sys, a system that enables controlled spatial sharing of
native CUDA compute and Vulkan graphics on NVIDIA GPUs.
An analysis of GPU scheduling, address translation, and workload dispatch
reveals how to coordinate the two execution stacks without replacing
either one.
\sys redirects CUDA channels into Vulkan's scheduling domain while
preserving the runtimes' separate data address spaces.
It then uses GPU front-end controls to partition execution resources
between the graphics and compute pipelines, with allocations adjustable
at runtime.
Together, these mechanisms let applications control resource sharing
without modifying GPU drivers or rewriting kernels and shaders.

We evaluate \sys across four application scenarios on three NVIDIA GPU
platforms, covering throughput and latency objectives for graphics,
compute, or both.
In embodied AI simulation, enabling concurrency improves throughput by up
to $1.62\times$ over the same asynchronous pipeline under default time
sharing.
In driving perception, partitioned co-execution reduces the measured
inference-budget miss rate from 80.1\% to zero while sustaining 60-FPS
rendering.
\end{abstract}

\settopmatter{printfolios=true}
\maketitle
{\renewcommand{\thefootnote}{}\footnotetext{\textsuperscript{*}Corresponding author: Jinyu Gu (\texttt{gujinyu@sjtu.edu.cn}).}}

\section{Introduction}
\label{sec:intro}

\begingroup
\setlength{\emergencystretch}{3em}
GPUs have traditionally served two largely separate domains---graphics
rendering and general-purpose computation---with individual GPUs typically
dedicated to one or the other.
Increasingly, modern applications require a single GPU to support both.
AI-enabled games render scenes while running models that control NPCs or
generate game
content~\cite{singh2026pubgally,nvidia2025acecharacters,goodai2024local,krafton_ai_usecases},
whether on a local GPU or a shared cloud
host~\cite{zhang2022pilotfish,chen2023combo};
VR/XR systems run real-time perception tasks beside immersive
rendering~\cite{huzaifa2021illixr,pan2024crisp,zhang2024boxr,chaudhary2019ritnet};
and simulators for embodied AI pair physics simulation with image
rendering~\cite{taomaniskill3,mittal2025isaaclab,rosenzweig2024batchrendering,zakka2025playground}.
Neural
rendering~\cite{vaidyanathan2023neuraltextures,zeltner2024neuralappearance,mueller2021radiancecache,chaitanya2017denoising},
on-device AI assistants~\cite{nvidia_gassist,nvidia2025chatrtx,yang2025ferretuilite},
and in-vehicle
systems~\cite{nvidia2022drivethor,masola2020adasgraphics,nvidia2024drivevulkansc}
mix the two as well.
\par\endgroup

These graphics--compute pairs carry different performance objectives: some
applications prioritize end-to-end throughput, which depends on both
sides, whereas others require one or both sides to meet explicit
service-level objectives (SLOs), such as a frame deadline or an
inference latency bound.
The system problem is to run graphics and compute concurrently, subject
to their dependencies, at high GPU utilization while meeting these
objectives.

\begin{figure}[!t]
  \centering
  \includegraphics[width=\columnwidth]{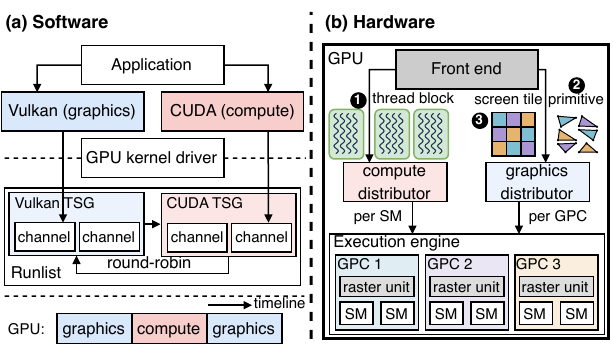}
  \caption{Challenges in graphics--compute sharing:
    (a)~separate TSGs serialize execution;
    (b)~compute dispatches to SMs, while graphics distributes work across GPCs.}
  \label{fig:walls}
\end{figure}

GPU sharing has been extensively studied for compute workloads, with
partitioning~\cite{bakita2023libsmctrl,DBLP:conf/asplos/0001XC0L026}
and scheduling~\cite{280914,DBLP:conf/asplos/ZhaoJP25,strati2024orion}
mechanisms that do not directly control native graphics.
Most existing graphics--compute approaches either support only time
sharing~\cite{zhang2025krypton} or do not support the widely used
rasterization pipeline~\cite{chen2023combo}.
Even CUDA in Graphics (CiG)~\cite{nvidia_cuda_cig_api}, the closest match to our requirements,
lacks fine-grained control over resource allocation between graphics
and compute~\cite{nvidia_tensorrtrtx_cig,nvidia_nvigi_gpusched}.
Table~\ref{tab:sharing-methods} summarizes these approaches.
Combining native graphics--CUDA concurrency with fine-grained,
runtime-adjustable resource allocation on real GPUs therefore remains
challenging.

Efficient graphics--compute sharing faces two challenges.

\noindent\textbf{Software stack.}
Compute and graphics workloads typically use different GPU interfaces,
such as CUDA for computation and Vulkan, OpenGL, or DirectX for rendering.\footnote{For
brevity, we use Vulkan to represent the graphics stack in this paper.}
Their runtimes create separate GPU contexts even within the same host
process.
On the NVIDIA GPUs targeted in this paper, the drivers place
CUDA and Vulkan command channels in separate \emph{timeslice groups}
(TSGs), as shown in Figure~\ref{fig:walls}(a).
The GPU schedules these groups in round-robin time slices, preventing the
two workloads from executing concurrently on the graphics engine.
Moving computation into the graphics execution stack requires either
rewriting CUDA kernels as Vulkan compute shaders or precompiling CUDA
kernels to PTX and launching them through a Vulkan
extension~\cite{vknvcudakernellaunch}.
Neither approach directly supports unmodified closed-source CUDA
libraries~\cite{nvidia_cublas_docs,nvidia_cudnn_docs}, nor does it preserve
the full range of advanced CUDA
features~\cite{nvidia_cuda_graphs,nvidia_cuda_cluster_launch,nvidia_cuda_dsm}.

\noindent\textbf{Hardware architecture.}
Even when compute and graphics can execute concurrently, their resource
sharing remains difficult to control.
For both Vulkan compute shaders and PTX kernels launched through the
Vulkan extension, the GPU's default simultaneous compute and graphics
(SCG) mode subjects graphics and compute work to best-effort arbitration
over shared SM resources to improve utilization~\cite{dash2022scgpatent}.
Meeting each workload's performance objectives instead requires flexible
resource allocation that limits interference.

Compute and graphics differ in two architectural aspects
(Figure~\ref{fig:walls}(b)).
First, their \emph{dispatch paths} differ: compute kernels reach SMs via
the compute work distributor, whereas graphics shaders use the graphics
pipeline.
Second, their \emph{dispatch granularities} differ at both the workload
and hardware levels.
Compute work is decomposed into thread blocks, whereas graphics work is
organized into primitive batches for vertex processing and screen regions
for pixel processing.
These units also target different hardware levels: compute blocks are
dispatched to individual SMs, whereas graphics work is first distributed
across graphics processing clusters (GPCs), owing to its dependence on
local fixed-function units~\cite{palmer2019scgpatent,dash2022scgpatent}.
These differences complicate fine-grained resource allocation across the
two pipelines.
Many isolation mechanisms, such as SM
masks~\cite{bakita2023libsmctrl,DBLP:conf/asplos/0001XC0L026}, act on only one
dispatch path and cannot coordinate resource allocation across the two
pipelines.

We bridge these software and hardware barriers with two mechanisms.
First, inspired by MPS's use of a shared TSG for multiple CUDA contexts,
we extend this scheduling arrangement across CUDA and Vulkan.
We redirect CUDA channels into Vulkan's TSG while preserving separate
address spaces, enabling concurrent execution across the two native
runtimes.
Second, we identify a common point of resource control in the GPU front
end, before work enters the two dispatch paths~\cite{palmer2019scgpatent},
and uncover an undocumented interface that switches SCG into a spatially
partitioned mode.
We use this interface to partition TPCs (groups of SMs) between graphics
and compute while preserving each pipeline's own dispatch granularity.
We make this allocation adjustable at runtime while keeping graphics
resources as balanced as possible across GPCs.

This paper presents \sys (CUDA + Vulkan), a system that combines these
mechanisms to enable concurrent graphics and compute execution with
controllable spatial sharing.
\sys preserves the CUDA execution stack and provides lightweight
partitioning to limit interference between the two workloads.

We evaluate \sys across embodied AI simulation, driving perception,
AI gaming, and VR on three NVIDIA GPU platforms.
These scenarios cover throughput objectives and SLOs for compute,
graphics, or both.
Redirection alone improves simulation throughput by up to $1.62\times$
over the same asynchronous pipeline under default time sharing.
Partitioning enables SLO attainment in tested configurations where
concurrency alone is insufficient.
For example, in driving perception, \sys reduces the measured fraction
of inferences exceeding their latency budget from 80.1\% to zero while
sustaining 60-FPS rendering.

This paper makes the following contributions:
\begin{itemize}[nosep,leftmargin=*]
  \item We demystify the software and hardware mechanisms underlying
        concurrent graphics and compute execution on NVIDIA GPUs.
  \item We develop a mechanism that enables spatial sharing across CUDA
        and Vulkan runtimes while preserving their separate address spaces.
  \item We design a lightweight partitioning mechanism that provides
        performance isolation between graphics and compute workloads.
  \item We evaluate \sys across four application scenarios on three GPU
        platforms, demonstrating higher simulation throughput and
        improved SLO attainment for graphics and compute.
\end{itemize}

\begin{figure}[b]
  \centering
  \includegraphics[width=\columnwidth]{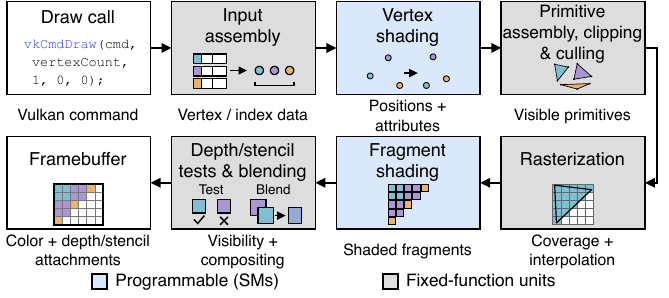}
  \caption{A simplified rasterization pipeline: programmable stages run on
    SMs (blue), while fixed-function stages use dedicated hardware (gray).}
  \label{fig:basics:pipeline}
\end{figure}

\section{Background and Motivation}
\label{sec:motivation}

\subsection{Rasterization Pipeline}
\label{sec:motivation:pipeline}
We focus on rasterization, the mainstream graphics pipeline; standalone
ray-tracing dispatches follow the compute path on NVIDIA
GPUs~\cite{nvidia_nsight_compute_warps}.
Rasterization converts geometry into images through vertex processing,
primitive assembly and clipping, rasterization, fragment processing, and
testing and blending (Figure~\ref{fig:basics:pipeline}).
\emph{Vertex shading} computes positions and attributes;
\emph{fragment shading} computes material and lighting effects.
Both run on SMs, while fixed-function stages use dedicated
hardware~\cite{khronos_vulkan_pipelines,nvidia_ada_architecture}.
A \emph{draw call} initiates the pipeline~\cite{khronos_vkcmddraw},
which manages shader invocations and inter-stage dependencies.
Individual shader stages are not independently submitted and scheduled
like CUDA kernels.

\subsection{GPU Submission and Execution}
\label{sec:motivation:submission}
\begin{figure}[t]
  \centering
  \includegraphics[width=\columnwidth]{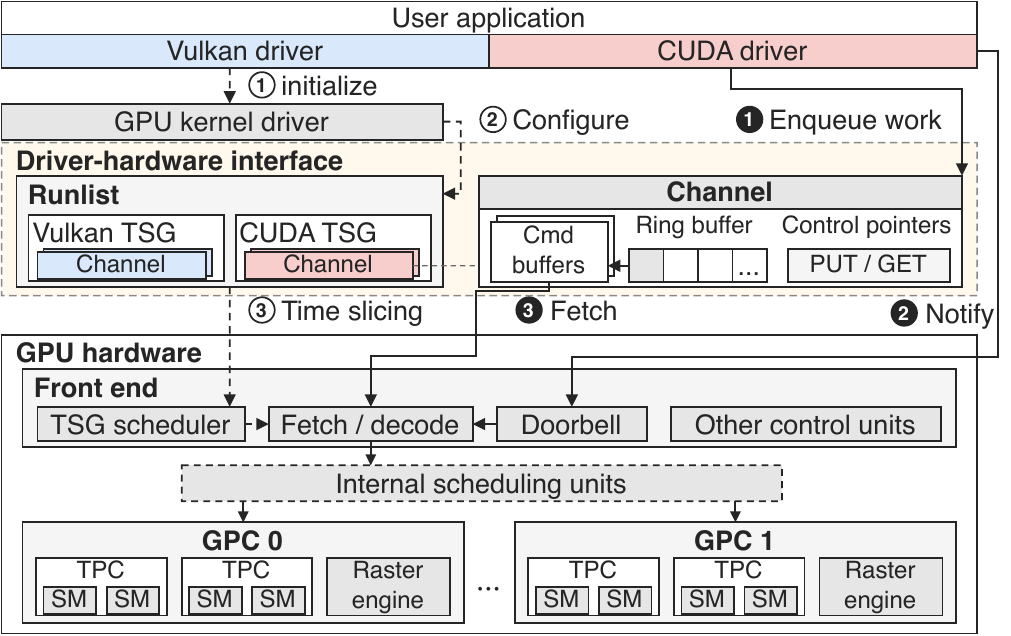}
  \caption{GPU submission and execution.
    Hollow circles mark \protect\whitecircled{1}~initialization,
    \protect\whitecircled{2}~configuration of the runlist--TSG--channel
    hierarchy, and \protect\whitecircled{3}~TSG time slicing.
    Filled circles mark \protect\circled{1}~command enqueueing,
    \protect\circled{2}~GPU notification, and
    \protect\circled{3}~command fetching.
    Execution resources are organized into GPCs, TPCs, and SMs;
    unit counts are schematic.}
  \label{fig:gpu-work-submit}
\end{figure}

Figure~\ref{fig:gpu-work-submit} traces GPU work from submission to
execution.
CUDA streams and Vulkan queues expose different programming interfaces,
but their user-mode drivers submit work through a common low-level
mechanism, the \emph{channel}.
A channel maintains a \emph{ring buffer} of entries referencing GPU
command buffers, together with \emph{control pointers} that track the
production and consumption of these entries.

During initialization, the user-mode driver requests channels through
the GPU kernel driver (\whitecircled{1}).
The kernel driver binds channels to \emph{timeslice groups} (TSGs) and
records runnable TSGs and their channels in a \emph{runlist}, establishing
the scheduling hierarchy visible to the GPU (\whitecircled{2}).
For the graphics/compute engine considered here, the hardware scheduler
time-slices between TSGs (\whitecircled{3}).
Runlists, TSGs, and channels form a driver--hardware interface: the
driver manages their objects and hardware-visible state, while the GPU
provides the corresponding scheduling and command-fetch
mechanisms~\cite{bakita2024demystifying}.

To submit work, the user-mode driver writes commands into GPU-accessible
memory, appends references to the channel's ring buffer, and advances
the producer control pointer (\circled{1}).
It then writes to a \emph{doorbell} to notify the GPU of pending work
on that channel (\circled{2}).
When the channel's TSG is scheduled, the GPU front end fetches and decodes
pending commands from its member channels (\circled{3}), making 
work from multiple channels eligible for concurrent
execution~\cite{bakita2024demystifying}.

Decoded graphics and compute commands feed internal scheduling units
that distribute work to the GPU's execution resources.
These resources form a hierarchy with three levels.
At the top level, the GPU consists of \emph{graphics processing clusters}
(GPCs).
Each GPC contains \emph{texture processing clusters} (TPCs), which in turn
contain \emph{streaming multiprocessors} (SMs) at the lowest level.
SMs execute CUDA kernels and programmable graphics shaders using
registers, arithmetic units, shared memory, and L1 cache.
Graphics processing additionally uses fixed-function units at the TPC
and GPC levels, including the rasterization engine within each GPC
shown in Figure~\ref{fig:gpu-work-submit}~\cite{nvidia_ada_architecture}.

The runtimes also create \emph{GPU contexts} that maintain execution
state and virtual address spaces.
In the default configurations examined here, CUDA and Vulkan bind their
compute and graphics channels to separate TSGs, even within one host
process.
Consequently, CUDA kernels and graphics shaders execute in separate
time slices on the GPU.

\begin{table}[b]
  \centering
  \small
  \setlength{\tabcolsep}{2.5pt}
  \renewcommand{\arraystretch}{1.1}
  \begin{tabular*}{\columnwidth}{@{\extracolsep{\fill}}lcccc@{}}
    \toprule
    Method & \makecell{Native\\Raster.} & \makecell{G/C\\Sharing}
    & \makecell{Real\\GPU}
    & \makecell{FG\\control} \\
    \midrule
    Compute-only & \no & --- & \yes & \partialmark \\
    KRYPTON & \yes & Temporal & \yes & \no \\
    Combo & \no & Spatial & \yes & \partialmark \\
    CRISP & \yes & Spatial & \no & \yes \\
    CiG & \yes & Spatial & \yes & \no \\
    \midrule
    \textbf{\sys} & \yes & Spatial & \yes & \yes \\
    \bottomrule
  \end{tabular*}
  \caption{GPU sharing capabilities.
    Native Raster.: native rasterization support.
    G: graphics; C: CUDA compute; FG: fine-grained.
    ---: no direct G/C sharing support.
    \yes\ yes, \no\ no, \partialmark\ limited to compute workloads
    or transformed rendering (Combo). \sys lists target capabilities.}
  \label{tab:sharing-methods}
\end{table}

\subsection{Limitations of Existing GPU Sharing Methods}
\label{sec:motivation:limitations}

Table~\ref{tab:sharing-methods} summarizes existing methods.
MPS and CUDA green contexts manage CUDA concurrency and resource
allocation~\cite{nvidia_mps,nvidia_green_contexts}.
SM/TPC masks, as used by libsmctrl, restrict thread-block dispatch through
kernel launch descriptors; disjoint permitted SM sets form spatial
partitions~\cite{bakita2023libsmctrl}.
Tally, MorphX, and REEF use kernel transformation, slicing, preemption,
or software scheduling to manage compute
execution~\cite{DBLP:conf/asplos/ZhaoJP25,ren2026morphable,280914}.
These compute-oriented methods do not directly partition resources
between native graphics and CUDA.
Among graphics--compute systems, KRYPTON's evaluated Vulkan path uses
time sharing~\cite{zhang2025krypton}; Combo enables spatial sharing but
requires transformed OptiX/CUDA rendering~\cite{chen2023combo}; and CRISP
studies SM and cache partitioning in a simulator rather than providing
real-GPU controls~\cite{pan2024crisp}.
CUDA in Graphics (CiG) supports co-execution on supported platforms, but
its public interfaces offer priority hints rather than explicit SM/TPC
quotas~\cite{nvidia_nvigi_gpusched,nvidia_tensorrtrtx_cig}.
Graphics-capable MIG configurations on selected Blackwell products
provide hardware isolation, but restrict partitioning to predefined
profiles~\cite{nvidia_mig_profiles}.
Our goal is to combine native graphics--CUDA spatial coexecution with
runtime-adjustable, fine-grained resource allocation on real GPUs.

\section{The \sys Framework}
\label{sec:overview}

\subsection{Design Goals}
\label{sec:overview:goals}

The submission boundary and resource-control limitations described in
\S\ref{sec:motivation} motivate three goals for \sys.
First, enable spatial sharing between Vulkan graphics and CUDA compute
workloads while preserving their native execution stacks, without
modifying the GPU driver or rewriting CUDA kernels or Vulkan shaders.
Second, provide fine-grained resource allocation between graphics and
compute, adjustable at runtime to support throughput and SLO objectives.
Third, support practical integration through interfaces that let
applications control sharing and resource allocation using existing
command-submission abstractions, such as CUDA streams and Vulkan queues,
within and across processes.

\subsection{Architecture and Workflow}
\label{sec:overview:architecture}

\begin{figure}[t]
  \centering
  \includegraphics[width=\columnwidth]{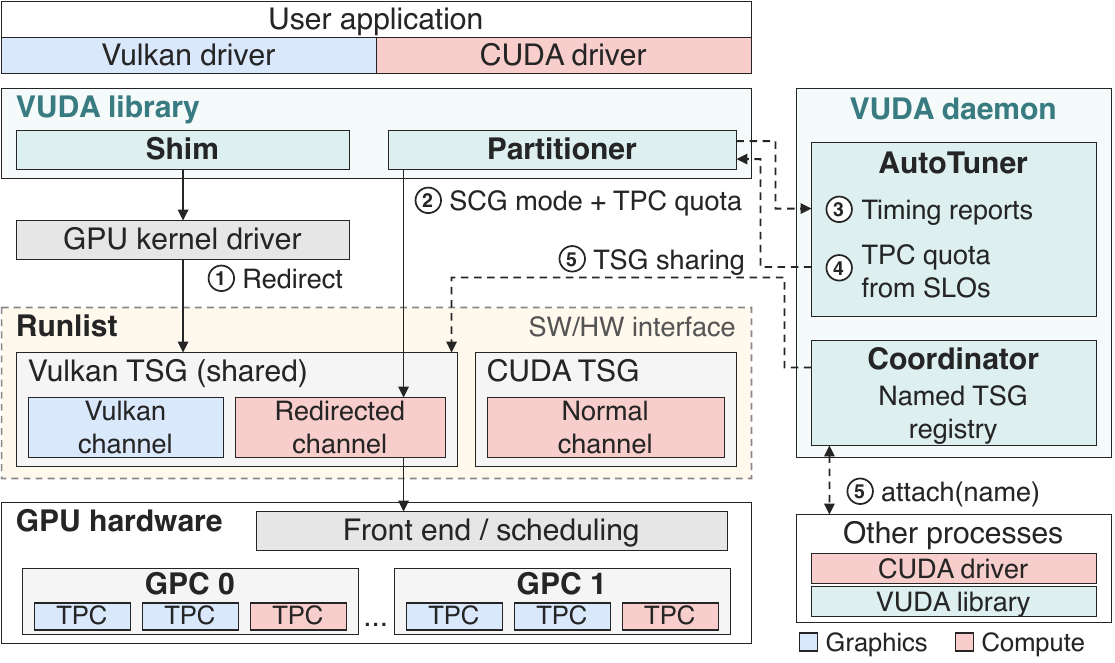}
  \caption{\sys architecture and workflow.}
  \label{fig:arch}
\end{figure}

\sys meets these goals by placing selected CUDA channels in the
graphics runtime's TSG and controlling the TPC allocation between the
concurrently executing workloads.
Figure~\ref{fig:arch} shows its three components: a
\emph{shim}, a \emph{partitioner}, and a \emph{daemon}.
The shim and partitioner form one library preloaded into the application
process; the daemon is a system service shared by applications using
\sys on the GPU.

\myparagraph{Shim.}
The shim intercepts CUDA's channel-creation calls to the driver.
It places the channels reserved through \sys's interface
(\S\ref{sec:overview:interface}) into the graphics runtime's TSG
(\whitecircled{1} in Figure~\ref{fig:arch}).
Each redirected channel retains CUDA's address space, allowing CUDA
kernels to execute unchanged.
\S\ref{sec:software} explains how the shim establishes this shared
scheduling arrangement and preserves correct address translation for
both workload data and shared control state.

\myparagraph{Partitioner.}
The partitioner controls how graphics and compute divide the GPU's TPCs.
It uses simultaneous compute and graphics (SCG) controls to select
\emph{intra-SM} co-residency, where both workloads can occupy the same
SMs, or \emph{inter-SM} partitioning, where they use separate TPCs.
In the latter mode, it also sets a compute TPC quota.
These settings are submitted as GPU control commands
(\whitecircled{2}) and take effect when the GPU processes them.
\S\ref{sec:hardware:scg} explains how the controls coordinate the two
dispatch paths, and \S\ref{sec:hardware:balance} describes how requested
quotas are aligned to balanced allocations across GPCs.
The partitioner also collects per-frame and per-kernel timings for the
tuner.

\myparagraph{Daemon.}
The daemon has two roles: tuning the partition and coordinating sharing
across processes.
The \emph{tuner} receives timing events from the partitioner
(\whitecircled{3}) and evaluates them against user-specified SLOs:
a frame-time target for graphics and a latency bound or
throughput floor for compute.
It returns the next quota to the partitioner (\whitecircled{4}).

The \emph{coordinator} extends sharing across processes (\whitecircled{5}).
A graphics application registers its TSG with the daemon under a name.
A CUDA application attaches to that name, and the shim places its
reserved channels into the registered TSG.
The address-space organization that permits channels from different
clients to share a TSG is explained in
\S\ref{sec:software:anatomy}.

\subsection{Programming Interface}
\label{sec:overview:interface}

\begin{figure}[tb]
  \centering
  \includegraphics[width=\columnwidth]{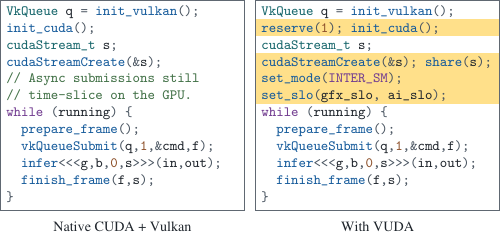}
  \caption{Native and \sys application code; shaded lines contain added calls.}
  \label{fig:programming-interface}
\end{figure}

Figure~\ref{fig:programming-interface} shows an application that combines
Vulkan rendering with CUDA inference using the interfaces in
Table~\ref{tab:vuda-interface}.
The native version submits both workloads without an intervening wait,
yet the GPU time-slices between their TSGs.
The shaded \sys calls enable spatial sharing while keeping the submission
loop unchanged; helpers abbreviate setup and frame completion.

\begin{table}[t]
  \centering
  \footnotesize
  \captionsetup{skip=4pt}
  \caption{\sys's programming interface.}
  \label{tab:vuda-interface}
  \setlength{\tabcolsep}{3pt}
  \renewcommand{\arraystretch}{1.0}
  \begin{tabular}{@{}>{\raggedright\arraybackslash}p{0.26\columnwidth}%
    >{\raggedright\arraybackslash}p{\dimexpr0.74\columnwidth-2\tabcolsep\relax}@{}}
    \toprule
    \textbf{Interface} & \textbf{Description} \\
    \midrule
    \code{reserve(n)}
      & Reserve $n$ channels before CUDA context creation. \\
    \code{share(stream)} /\newline \code{unshare(stream)}
      & Bind a stream to a reserved channel for sharing, or return it to an ordinary channel. \\
    \code{attach(name)}
      & Attach to another process's named graphics TSG. \\
    \code{set\_mode(mode)}
      & Select \mbox{\code{INTRA\_SM}} or \mbox{\code{INTER\_SM}} sharing. \\
    \code{set\_quota(q)}
      & Request $q$ compute TPCs; return the balanced quota. \\
    \code{set\_slo}\newline \code{(gfx, compute)}
      & Set graphics frame-time and compute latency or throughput SLOs for automatic quota tuning. \\
    \bottomrule
  \end{tabular}
\end{table}

The application initializes Vulkan, then calls \code{reserve(1)} before
CUDA context creation. The shim redirects the reserved channel into the
graphics TSG (\whitecircled{1} in Figure~\ref{fig:arch}).
After stream creation, \code{share(s)} binds the inference stream to that
channel; \code{unshare(s)} can later return it to an ordinary channel.

Next, \mbox{\code{set\_mode(INTER\_SM)}} selects TPC partitioning, and
\code{set\_slo} supplies a frame-time target (\code{gfx\_slo}) and an
inference throughput floor (\code{ai\_slo}) for automatic quota tuning.
The partitioner applies mode and quota settings through GPU control
commands (\whitecircled{2}), reports timings to the tuner
(\whitecircled{3}), and receives updated quotas in return
(\whitecircled{4}).
For manual allocation, applications can instead request a compute quota
with \mbox{\code{set\_quota(q)}}.

For cross-process sharing, the renderer registers its TSG under a name,
and the inference process calls \code{attach(name)} before creating its
reserved channels. The coordinator then connects those channels to the
registered TSG (\whitecircled{5}).
\S\ref{sec:software} and \S\ref{sec:hardware} detail the channel-sharing
and resource-control mechanisms, respectively;
\S\ref{sec:impl} describes their implementation.

\section{Sharing Across Software Boundary}
\label{sec:software}

The shim introduced in \S\ref{sec:overview:architecture} must place
selected CUDA channels in the graphics TSG while preserving the two
runtimes' data address spaces.
This requires correct address translation for both workload data and
the internal buffers used to process GPU commands.
We first explain how one TSG can support multiple address spaces, then
show how the shim handles data-plane and control-plane accesses.

\subsection{Anatomy: Two Contexts, One TSG}
\label{sec:software:anatomy}

\begin{figure}[t]
  \centering
  \includegraphics[width=\columnwidth]{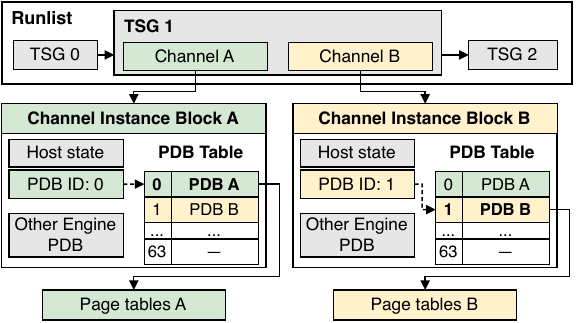}
  \caption{GPU submission and address-space organization.
    Channels in one TSG carry matching PDB tables in their instance
    blocks. Each channel's PDB ID selects its address space, allowing
    channels A and B to use different page tables within one TSG.}
  \label{fig:software:runlist-pdb}
\end{figure}

We investigate the submission structures by intercepting the runtimes'
\code{ioctl()} calls, inspecting the memory regions involved, and
comparing the observed layouts with public driver and hardware
definitions.
This analysis reconstructs the relationships from the runlist to TSGs,
channels, and their address-translation state
(Figure~\ref{fig:software:runlist-pdb}).
It reveals that channels in the same TSG can use different page tables.

Each channel has a \emph{channel instance block} in GPU memory that
stores its control and address-translation state.
A \emph{page-directory base} (PDB) identifies the root of a GPU page
table.
The instance block records a PDB for other engines, such as copy and
video engines, and a separate \emph{PDB table} for the graphics/compute
engine.
This table contains the page-directory bases of multiple subcontexts.
The channel's subcontext identifier, denoted \emph{PDB ID} in the figure,
selects an entry in that table.
The hardware can therefore select a channel's address space independently
of the TSG that determines when the channel is scheduled.

In the ordinary CUDA and Vulkan configurations we examined, a TSG
typically has one populated PDB entry, and its channels use PDB ID~0.
Our analysis of MPS shows how the same structure supports multiple
address spaces: channels from different CUDA clients join the server's
TSG, and each client's page-directory base occupies a distinct entry.
All channel instance blocks in the TSG carry the same table contents,
while channels from different clients select different PDB IDs.
Thus, one TSG can schedule channels from multiple contexts without
requiring their workload data to share a single address space.
\sys applies this arrangement to CUDA and Vulkan.

\subsection{Data Plane and Control Plane}
\label{sec:software:design}

\begin{figure}[t]
  \centering
  \includegraphics[width=\columnwidth]{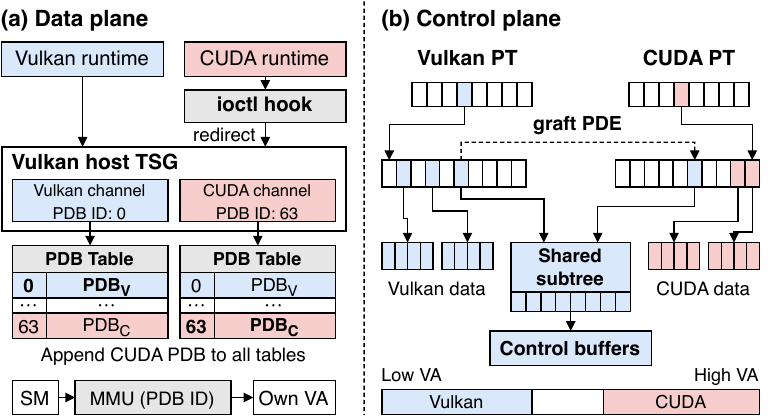}
  \caption{\sys's software design.
    (a)~Channel redirection enables concurrent execution with separate
    data address spaces.
    (b)~Page-table grafting makes shared control buffers accessible
    through both address spaces.}
  \label{fig:software:design}
\end{figure}

Putting the channels in one TSG removes the scheduling boundary, but it
does not by itself preserve correct memory access.
We distinguish two kinds of access by their source.
\emph{Data-plane} accesses are issued by the SMs as they execute CUDA
kernels and graphics shaders.
\emph{Control-plane} accesses are issued by units that process commands
and maintain execution state, including the GPU front end and command
macro processors.
\sys handles their address translation separately, as illustrated in
Figure~\ref{fig:software:design}.

\myparagraph{Data plane: redirecting CUDA channels.}
Either runtime's TSG could appear to be a candidate for hosting both
workloads.
Our traces show, however, that the Vulkan runtime performs extensive
graphics-specific initialization.
Using its existing TSG preserves this initialized graphics state and
avoids reconstructing it in a CUDA TSG.
\sys therefore selects the Vulkan TSG as the host and redirects the
reserved CUDA compute channels into it.

\sys intercepts CUDA's channel-creation \code{ioctl()} calls for these
reserved channels and changes their TSG binding to the selected Vulkan TSG.
The CUDA context retains its original page tables.
Its page-directory base is added at an unused index in the host TSG's
PDB tables, and its channels receive the corresponding PDB ID.
The tables remain consistent across all channels in the combined TSG;
Vulkan channels retain the PDB IDs for their existing address spaces.
Channels belonging to the same CUDA context select the same added entry.

During execution, the GPU MMU uses the subcontext identifier associated
with an SM memory request to select the appropriate PDB.
CUDA kernels therefore continue to access their data through CUDA's
page tables, while graphics shaders use Vulkan's page tables.
This preserves each workload's virtual addresses and data mappings while
allowing the two runtimes to submit work for concurrent execution.
The change occurs at channel setup and does not require translating
CUDA kernels into Vulkan compute shaders.

\myparagraph{Control plane: mapping shared control buffers.}
Correct SM accesses are necessary but insufficient.
We find that some control state is shared across the TSG, and accesses
to this state do not always use the issuing channel's PDB.
A control buffer created by Vulkan can consequently be accessed through
the CUDA subcontext's page tables.
Redirecting channels alone leaves this buffer unmapped in that address
space and can cause a GPU memory fault.

The runtimes' different allocation conventions provide a useful opportunity.
CUDA uses unified virtual addressing (UVA) for host and device
allocations~\cite{cuda_programming_guide}.
In the configurations we examined, CUDA allocations occupy high virtual
addresses, while Vulkan's internal allocator places the relevant control
buffers at low addresses.
This natural separation leaves room for Vulkan's control state in CUDA's
address space without relocating either runtime's data.

\sys exploits this layout by grafting the Vulkan
\emph{page-directory entries} (PDEs) covering the control buffers into
unused entries in CUDA's page tables (Figure~\ref{fig:software:design}(b)).
Reusing the existing subtrees preserves the buffers' mappings and makes
them accessible through either PDB.

With the shim establishing co-execution, the partitioner must control
how the workloads share execution resources.
\S\ref{sec:hardware} explains this second mechanism.

\section{Partitioning Across Hardware Pipeline}
\label{sec:hardware}

The shim in \S\ref{sec:software} enables CUDA--Vulkan concurrency.
The partitioner must then control the resource allocation to support
the performance goals of \S\ref{sec:overview:goals}, since interference
can still prevent either workload from meeting its SLO.
Compute and graphics reach the SMs through different dispatch paths, so
a mask that restricts compute placement does not directly restrict
graphics.
We explain these paths, introduce spatial partitioning through SCG
controls, and describe how \sys balances the allocation across GPCs.

\subsection{Anatomy: Two Roads to the SM}
\label{sec:hardware:anatomy}

\begin{figure}[t]
  \centering
  \includegraphics[width=\columnwidth]{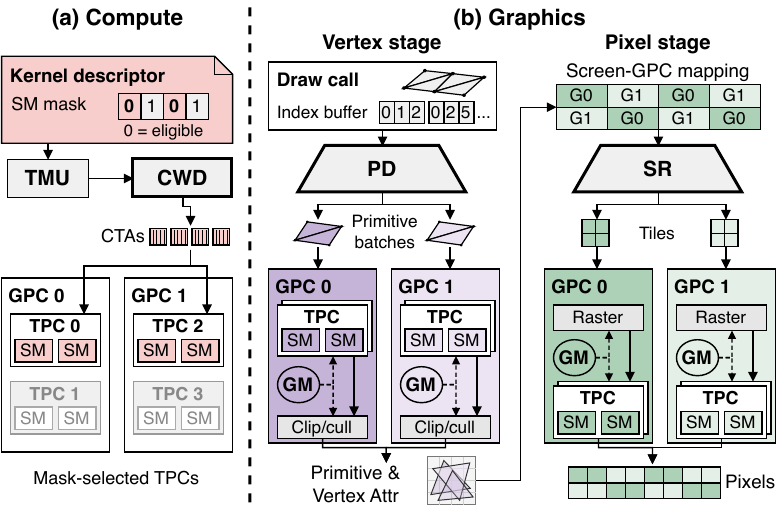}
  \caption{Compute and graphics dispatch paths.
    Launch masks constrain compute; graphics distributes primitive
    batches and screen tiles across GPCs.
    Solid arrows: data flow; dashed arrows: scheduling.}
  \label{fig:hardware:dispatch}
\end{figure}

We inspect the task metadata and command streams produced by CUDA kernel
launches and Vulkan queue submissions, and relate our observations to
NVIDIA's published patent
descriptions~\cite{palmer2019scgpatent,dash2022scgpatent}.
Figure~\ref{fig:hardware:dispatch} summarizes the resulting dispatch
model.
For graphics, we distinguish \emph{vertex processing} from
\emph{pixel processing}.
These terms denote portions of the pipeline that include both shader
execution and fixed-function operations; they do not refer only to the
vertex and fragment shaders.

\myparagraph{Compute dispatch.}
The task management unit passes a compute task to the
\emph{compute work distributor} (CWD).
It distributes the task's cooperative thread arrays (CTAs), or thread
blocks, to TPCs according to scheduling priority and available capacity.
It also honors the SM/TPC mask in the task descriptor: each CTA may
execute only on an allowed SM.
The descriptor thus specifies the eligible resources, while the
distributor selects where to place each block within that set.
Disjoint masks give compute workloads separate execution resources, as
used by libsmctrl~\cite{bakita2023libsmctrl}, but do not exclude work
arriving through the graphics path.

\myparagraph{Graphics dispatch: vertex processing.}
The \emph{primitive distributor} (PD) sends batches of graphics
primitives to GPCs.
For a draw using an indexed list of triangles, each group of three
indices identifies the vertices of a triangle.
For example, $(0,1,2)$ and $(2,3,0)$ describe two primitives that share
vertices.
For $N$ primitives spread approximately evenly across $M$ GPCs, each
processes roughly $N/M$ primitives, although equal counts need not imply
equal shader execution time.
Vertex processing produces the geometry and attributes needed by the
subsequent pixel-processing path.

\myparagraph{Graphics dispatch: pixel processing.}
The \emph{screen router} (SR) routes primitive data according to its
coverage of the output image.
Each screen region has an assigned GPC that handles its pixel
processing.
For example, a triangle covering regions assigned to two different GPCs
contributes pixel-processing work to both.
Its destination is determined by screen coverage, which can differ from
the placement chosen earlier for vertex processing.
Within a GPC, the \emph{GPC manager} (GM) coordinates work across
TPCs and fixed-function units, including the rasterization engine.
The SMs execute shader work as part of this local pipeline.

Our inspection of the command buffers for the examined draws found
no equivalent per-draw field for specifying a resource mask to the PD,
SR, or GM.
The compute descriptor's mask is consumed by a different distributor and
does not govern these decisions.
Restricting CUDA's eligible resources alone therefore leaves graphics
free to use overlapping resources.
A graphics--compute partition must coordinate both dispatch paths.

\subsection{Simultaneous Compute and Graphics}
\label{sec:hardware:scg}

Our measurements of the default configuration show \emph{intra-SM SCG}:
warps from CUDA kernels and graphics shaders can reside on the same SMs,
consistent with NVIDIA's patent description~\cite{dash2022scgpatent}.
The hardware admits work according to available resources and its
arbitration rules, using spare capacity without providing an explicit
split between workloads.
The capacity available to one workload consequently depends on the
other's demand.
For latency-sensitive rendering, this makes concurrency alone
insufficient: increasing compute activity can delay shader execution
even though both workloads remain runnable.

\begin{figure}[t]
  \centering
  \includegraphics[width=\columnwidth]{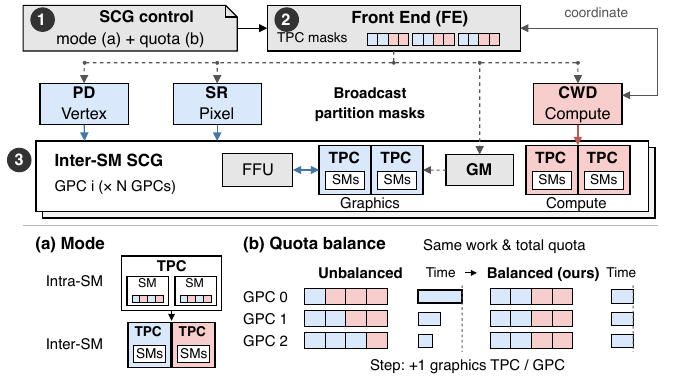}
  \caption{SCG mode control and partitioning across GPCs.
    The inset illustrates imbalance from uneven allocation.
    Blue: graphics; red: compute.}
  \label{fig:hardware:scg-partition}
\end{figure}

An earlier patent describes \emph{inter-SM SCG}, which assigns separate
TPCs to the two pipelines~\cite{palmer2019scgpatent}.
Its allocation unit is a TPC, including its constituent SMs.
This design provides a way to control how much execution capacity each
pipeline can use while they run concurrently.
Figure~\ref{fig:hardware:scg-partition} contrasts the two modes.

Guided by these descriptions, we identify an SCG control interface
exposed through GPU commands.
Our black-box tests show that these controls can switch between the two
modes on the hardware we examined and set a compute TPC quota in
inter-SM mode.
\sys uses this interface to enable partitioning and configure the
resource split.

In this mode, the \emph{front end} (FE) coordinates TPC ownership before
work reaches the separate dispatch paths.
It communicates the allocation to the compute and graphics dispatch
logic.
The CWD places compute blocks within the compute allocation, while the
graphics pipeline dispatches shader work within its own allocation.
The two paths keep their existing work-distribution mechanisms but
operate under a common resource assignment.
This establishes the boundary that a compute-only mask cannot provide.

\subsection{Avoiding Load Imbalance}
\label{sec:hardware:balance}

Two partitions with the same total graphics allocation can have different
completion times: the distribution of TPCs across GPCs also matters.
A partition can reduce a GPC's shading capacity without reducing the
graphics work assigned to it.

This mismatch arises from the locality of graphics processing
(\S\ref{sec:hardware:anatomy}).
Once rasterization begins, subsequent shader work depends on local
pipeline state, so spare graphics capacity in other GPCs cannot directly
take over.
With comparable work assigned to each GPC, a GPC left with fewer graphics
TPCs can therefore delay frame completion even when capacity remains
available elsewhere (Figure~\ref{fig:hardware:scg-partition}).

\sys therefore exposes partition levels that transfer one TPC per
active GPC between workloads at each step.
For $N_{\mathrm{GPC}}$ active GPCs, the total quota changes by
$N_{\mathrm{GPC}}$ TPCs per step.
These coordinated changes avoid reducing graphics capacity in only a
subset of the groups.
The range is limited by the enabled topology and the requirement to
retain at least one graphics TPC per active GPC during
rendering~\cite{palmer2019scgpatent}.

For example, the RTX~4090 configuration in our study has 11 active GPCs:
two contain five TPCs each, and nine contain six, for 64 TPCs in total.
A coordinated step therefore transfers 11 TPCs.
This policy limits imbalance introduced by partitioning; it does not
assume that primitive complexity or screen-space shading cost is uniform
for every scene.

\section{Implementation}
\label{sec:impl}

\sys implements channel redirection (\S\ref{sec:software}) in a
user-space library loaded through \code{LD\_PRELOAD}.
It intercepts the native \code{ioctl()} calls issued by the CUDA and
Vulkan user-mode drivers.
The shim uses the driver's existing object-duplication ioctl to obtain a
handle to the Vulkan TSG in CUDA's driver client, then redirects CUDA's
subcontext and channel allocations by rewriting their parent handles.
It preserves CUDA's virtual-address-space handle and restores the
modified input arguments before returning.
The driver performs the binding through its existing allocation path,
requiring neither a new ioctl interface nor a driver patch.
For SCG control (\S\ref{sec:hardware}), the partitioner maintains an
internal \emph{shadow queue} backed by a dedicated \emph{shadow channel}.
It submits the undocumented GPU control commands for SCG mode selection
and compute TPC quotas directly through this channel, leaving the
application's CUDA kernels and Vulkan shaders unchanged.

\section{Evaluation}
\label{sec:eval}

We evaluate how redirection enables graphics--compute sharing and how
partitioning changes its performance trade-offs.
Microbenchmarks isolate co-residency, quota sensitivity, and runtime
reconfiguration (\S\ref{sec:eval:micro}).
Four application scenarios then examine aggregate throughput
(\S\ref{sec:eval:sim}), a compute SLO (\S\ref{sec:eval:driving}),
a graphics SLO (\S\ref{sec:eval:gaming}), and joint graphics and
compute SLOs (\S\ref{sec:eval:vr}).
The results show when redirection alone suffices and when meeting an
application objective also requires an appropriate partition.

\subsection{Setup}
\label{sec:eval:setup}

\myparagraph{Platforms.}
We evaluate \sys on three NVIDIA GPUs (Table~\ref{tab:platforms}).
All experiments pair native Vulkan graphics with CUDA compute on the
same GPU under Linux.

\begin{table}[b]
  \centering
  \small
  \caption{Evaluated platforms. SM clocks are fixed on RTX~4090/A800
    and measured on Orin.}
  \label{tab:platforms}
  \setlength{\tabcolsep}{3pt}
  \begin{tabular*}{\columnwidth}{@{\extracolsep{\fill}}llclr@{}}
    \toprule
    GPU & Arch. & \makecell{TPCs /\\GPCs} & Driver & \makecell{SM\\(MHz)} \\
    \midrule
    RTX 4090 & Ada & 64 / 11 & 610.43.02 & 2100 \\
    A800-SXM4 & Ampere & 54 / 7 & 610.57.04 & 1410 \\
    Jetson AGX Orin & Ampere & 8 / 2 & 540.4.0 & 1300.5 \\
    \bottomrule
  \end{tabular*}
\end{table}

\myparagraph{Configurations.}
\emph{Baseline} time-slices CUDA and Vulkan in separate TSGs.
\emph{Redirect-only} joins them in one TSG using the default SCG mode
(inter-SM on GA100). \emph{VUDA-full} adds inter-SM partitioning with
a compute quota $q$, measured in TPCs.
Applications use fixed quotas selected from sweeps; microbenchmarks
also test runtime changes.

\myparagraph{Measurements.}
Unless stated otherwise, we report medians and min--max ranges across
runs. CDFs pool recorded samples, with bands spanning per-run CDFs.
Each scenario specifies its timing boundaries and SLO criteria.

\subsection{Microbenchmarks}
\label{sec:eval:micro}

On RTX~4090, we pair a 1080p Vulkan renderer with a continuously
running synthetic CUDA workload.
Vertex-bound and fragment-bound scenes each take approximately
25\,ms alone, exposing different graphics bottlenecks.
Each process measures its standalone rates and then 120 co-running
frames after six warm-up frames, with five repetitions per configuration.
Graphics frame rate and completed CUDA kernels/s are measured over
the same interval and normalized to their respective standalone rates.

\begin{figure}[t]
  \centering
  \includegraphics[width=\columnwidth]{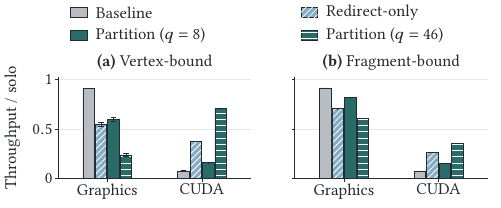}
  \caption{Scheduling ablation. Throughput is normalized to solo;
    bars and whiskers show medians and min--max across five runs.}
  \label{fig:ablation}
  \Description{Graphics and CUDA throughput under time sharing,
    redirect-only execution, and partitioning at quotas of 8 and 46 TPCs,
    for vertex-bound and fragment-bound scenes.}
\end{figure}

\myparagraph{Concurrency and partitioning.}
Default time sharing preserves 92\% of standalone graphics throughput
but leaves CUDA with only 8\% (Figure~\ref{fig:ablation}).
Redirect-only increases CUDA throughput by $4.9\times$ for the
vertex-bound scene and $3.5\times$ for the fragment-bound scene,
while graphics throughput falls to 55\% and 71\%, respectively.
Co-residency thus improves compute progress at graphics' expense.
Partitioning makes this trade-off adjustable by controlling the
allocation between the two workloads.
Relative to redirect-only, $q=8$ raises graphics throughput to 60\%
and 82\% of solo in the two scenes, respectively, but reduces CUDA
throughput.
The larger quota, $q=46$, favors CUDA, which reaches 71\% and 36\%
of solo, while graphics performance falls further.

\begin{figure}[tb]
  \centering
  \includegraphics[width=\columnwidth]{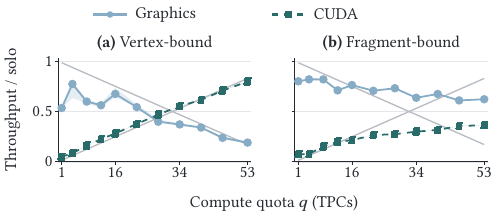}
  \caption{Throughput versus compute allocation.
    Medians and min--max bands cover five runs.
    Gray lines show the resource shares $q/64$ and $(64-q)/64$.}
  \label{fig:quota-sensitivity}
  \Description{Normalized graphics and CUDA throughput versus compute
    TPC quota for vertex-bound and fragment-bound scenes. Bands show the
    full range across runs; gray lines indicate resource shares.}
\end{figure}

\myparagraph{Quota sensitivity.}
Figure~\ref{fig:quota-sensitivity} sweeps 11 compute quotas from 1 to
53 TPCs to characterize the hardware mechanism.
Separate hardware checks observe the requested allocation at seven
tested quotas in both scenes.
With the vertex-bound scene, compute throughput rises from 5\% to 80\%
of its solo rate, while graphics performance varies non-monotonically.
The same allocation also produces different results across workloads:
at $q=53$, CUDA achieves 80\% of its solo rate alongside the
vertex-bound scene, but only 36\% alongside the fragment-bound scene.
TPC allocation alone therefore cannot predict throughput.
Across quotas, the median sum of normalized rates spans
0.59--0.99 and 0.88--1.01, respectively.
Partitioning offers flexibility in sharing resources, with a
workload-dependent cost to aggregate throughput.

\begin{figure}[t]
  \centering
  \includegraphics[width=\columnwidth]{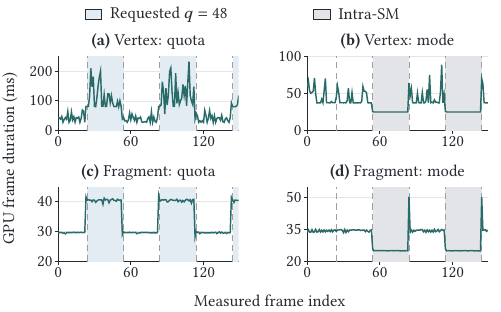}
  \caption{GPU frame-time response to quota changes
    ($8\leftrightarrow48$, left) and SCG-mode changes ($q=22$, right).
    Dashed lines mark requests; shading denotes $q=48$ or intra-SM mode.
    All frames are shown; vertical scales differ.}
  \label{fig:retune}
  \Description{Four frame-duration traces cover both scenes during
    quota and mode changes. All 600 samples are shown, including the
    232 millisecond peak in the vertex-bound quota trace.}
\end{figure}

\myparagraph{Runtime reconfiguration.}
Across three runs without contention, the median per-run mean host
cost of synchronous control is $71\,\mu$s.
Control commands included in rendering add $0.64$--$0.85\,\mu$s of
GPU time. With CUDA active, the host cost rises to $119$--$130\,\mu$s.
These measurements reissue unchanged settings and quantify control cost.
Figure~\ref{fig:retune} shows four 150-frame traces of actual
reconfiguration, with requests every 30 submitted frames.
Alternating between $q=8$ and $q=48$ produces recurring frame-time
plateaus near 29.7 and 40.5\,ms for the fragment-bound scene.
The vertex-bound scene exhibits large fluctuations, reaching
232\,ms, and mode changes also produce transient spikes.
Dynamic control must therefore account for workload-dependent
frame-time transients despite its low command cost.

\subsection{Throughput: Embodied AI Simulation}
\label{sec:eval:sim}

We evaluate four representative ManiSkill~3.0.1
scenes~\cite{taomaniskill3}, labeled A--D, on RTX~4090 and A800,
using PhysX~5.10 and SAPIEN~3's Vulkan renderer~\cite{Xiang_2020_SAPIEN}.
Scenes C/D have heavier rendering workloads per environment than A/B.
Baseline and \sys use the same seeded random actions and an
asynchronous pipeline whose images lag the simulated state by one
control step.
The baseline runs this pipeline with default time sharing;
\sys adds redirection without partitioning.
The comparison isolates the additional benefit of redirection within
this asynchronous pipeline.

\begin{figure}[t]
  \centering
  \includegraphics[width=\columnwidth]{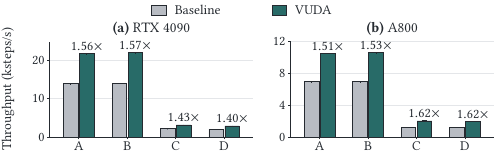}
  \caption{Simulation throughput. Bars show medians; whiskers show
    min--max ranges. Annotations give speedup over the asynchronous baseline.}
  \label{fig:sim}
  \Description{Throughput of four manipulation scenes on RTX 4090 and
    A800, comparing the same asynchronous simulation pipeline with
    default time sharing and VUDA redirection. All sixteen bars show
    medians and measured ranges. VUDA improves throughput in every
    scene without a compute partition.}
\end{figure}

\myparagraph{Throughput gains.}
Figure~\ref{fig:sim} reports completed environment steps per second,
including visual observations.
Redirection raises throughput by $1.40$--$1.57\times$ on RTX~4090
and $1.51$--$1.62\times$ on A800.
Peak throughput reaches 22.1 and 10.6 thousand steps/s, respectively.
Gains persist across all four scenes,
showing that redirection alone can improve aggregate throughput when
physics and rendering have independent work to overlap.
This experiment measures simulation throughput; it does not evaluate
the effect of delayed observations on policy learning.

\subsection{Compute SLO: Driving Perception}
\label{sec:eval:driving}

On Jetson AGX Orin, we pair SegFormer-B0 trained on Cityscapes
($1024\times1024$ inputs, TensorRT~10.3) with an offscreen Vulkan
renderer paced at 60\,Hz.
All configurations use the same inference engine and three 20-s runs
after a 3-s warm-up, with the GPU at 1.30\,GHz.
Inferences run continuously; their latency includes transfers but
excludes output validation. We compare it against a 33.33-ms budget
for 30-Hz perception.

\begin{figure}[t]
  \centering
  \includegraphics[width=\columnwidth]{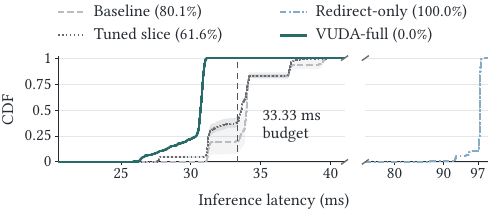}
  \caption{Driving-perception latency with 60-Hz rendering.
    CDFs pool three runs; bands show the range of per-run CDFs.
    Legend percentages are misses of the 33.33-ms budget.
    The axis break omits no samples; horizontal scales differ.}
  \label{fig:driving}
  \Description{Empirical inference-latency CDFs for baseline time sharing,
    tuned time sharing, redirect-only, and VUDA with a four-TPC compute
    quota. Their pooled miss rates are 80.1, 61.6, 100, and zero percent.
    Shading shows variation across three runs. All samples, including
    redirect-only's low tail, are retained.}
\end{figure}

\myparagraph{Meeting the compute SLO.}
After screening three shader loads, we report the intermediate load,
where the deadline distinguishes the configurations.
From seven priority/time-slice settings, we select the best inference
performance subject to preserving 60\,FPS: higher compute priority
and a 16-ms compute time slice.
VUDA-full uses four of eight TPCs for compute, selected from
quotas $q\in\{4,5,6\}$ at the same rendering load.
All four configurations sustain 60\,FPS in every run.
Figure~\ref{fig:driving} shows that baseline and tuned time sharing
miss the inference budget for 80.1\% and 61.6\% of samples,
respectively; redirect-only misses it for every sample.
With partitioning, all 1,861 measured inferences finish within the
budget: pooled p99 is 31.07\,ms and the maximum is 31.61\,ms.
Thus, redirection and partitioning together meet the compute SLO
while preserving the rendering rate.

\myparagraph{Load sensitivity.}
At a lighter 60-Hz rendering load, baseline already meets the budget.
Across four shader loads with unpaced rendering, $q=6$ lowers median
inference latency by 17.6--39.4\% while retaining 82.6--95.6\% of
baseline frame rate, but does not keep every inference within 33.33\,ms.
A separate unpaced ablation finds that partitioning without redirection
leaves median latency near baseline (69.0\,ms); combining them lowers
it to 56.8\,ms. The mechanisms are complementary, and SLO attainment
depends on the rendering demand.

\subsection{Graphics SLO: AI Gaming}
\label{sec:eval:gaming}

\begin{figure}[t]
  \centering
  \includegraphics[width=\columnwidth]{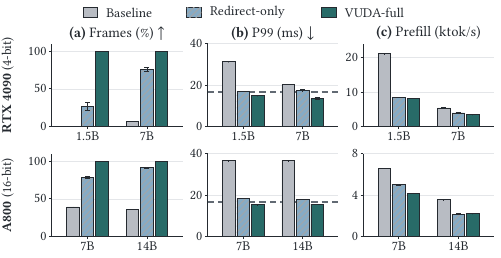}
  \caption{AI gaming. RTX~4090: pooled GPU-budget attainment and GPU-time
    p99; A800: median rendered/scheduled frames and release-to-completion
    p99. Prefill bars show medians; whiskers show min--max.
    Dashed lines mark 16.67\,ms.}
  \label{fig:gaming}
  \Description{Graphics performance and prefill throughput for baseline,
    redirect-only, and fixed partitioning. RTX 4090 uses 1.5B and 7B
    models with 4-bit quantized weights; A800 uses 7B and 14B models
    with 16-bit floating-point weights. The first column
    reports GPU-budget attainment on RTX 4090 and the fraction of
    scheduled frames rendered on A800. The second column reports
    GPU-time p99 and release-to-completion p99, respectively. Full
    partitioning puts both p99 measures below the 16.67-ms budget.}
\end{figure}

We model AI gaming with a 4K, nine-pass offscreen Vulkan renderer
paced at 60\,Hz and Qwen2.5 inference through unmodified llama.cpp.
The inference workload continuously prefills 2,048-token prompts.
RTX~4090 runs 1.5B and 7B models with 4-bit quantized weights
(Q4\_K\_M), using four and two concurrent inference streams,
respectively. A800 runs 7B and 14B models with 16-bit floating-point
weights (FP16), using four streams.
Each comparison fixes the GPU, model weights, and inference settings.
We report demanding graphics settings where rendering alone consumes
67--78\% of the 16.67-ms frame budget.
Figure~\ref{fig:gaming} compares baseline, redirect-only, and
VUDA-full with $q=8$ on RTX~4090 and $q=20$ on A800.
Quota sweeps select these settings to meet the graphics budget while
retaining inference throughput.

\myparagraph{Protecting graphics.}
Redirection improves graphics performance on both GPUs, but its p99
remains above the frame budget at these loads.
Partitioning lowers GPU-time p99 to 13.44--14.98\,ms on RTX~4090
and completion-latency p99 to 15.32--15.67\,ms on A800.
Every measured RTX~4090 frame stays within the GPU budget, and
A800 drops no frames with partitioning.
Thus, redirection and partitioning together protect graphics at
loads where redirection alone is insufficient.

\myparagraph{Inference throughput.}
The graphics benefit has a cost: VUDA-full retains 39--64\% of baseline
prefill throughput across these four configurations.
This cost depends on rendering demand. In a lighter RTX~4090
experiment, where solo rendering uses 45\% of the frame budget,
redirect-only already stays within the GPU budget; partitioning
further reduces GPU-time p99 below 9\,ms while retaining
91--98\% of baseline prefill throughput.

\subsection{Dual SLOs: VR Rendering and Perception}
\label{sec:eval:vr}

Our synthetic VR scenario on RTX~4090 combines a 90-Hz stereo Vulkan
renderer ($1440\times1600$ per eye) with a 60-Hz CUDA workload
representing perception.
Rendering has a 9-ms budget within each 11.11-ms frame interval;
perception has a 16.67-ms deadline.
Solo rendering takes 5.8\,ms; nominal and heavy perception take
6.4 and 8.1\,ms.
Each configuration uses five 12-s runs after a 1.5-s warm-up, with fixed
quotas. Latency includes queueing from scheduled release; miss rates
cover completed jobs.
Joint attainment requires at most 1\% rendering and 5\% perception
misses in every run; the 5\% limit was selected after inspecting the data.

\begin{figure}[tb]
  \centering
  \includegraphics[width=\columnwidth]{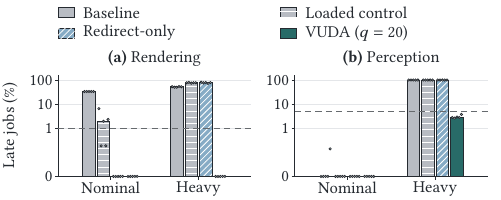}
  \caption{VR deadline misses. Bars show medians; each dot is one run.
    Loaded control retains redirection with partitioning disabled.
    Dashed lines mark the 1\% and 5\% limits.
    The vertical scale is linear below 0.1\% and logarithmic above.}
  \label{fig:vr}
  \Description{Rendering and perception miss rates at nominal and heavy
    loads, comparing time sharing, a loaded control, redirect-only, and
    VUDA with a 20-TPC compute quota. All five runs per configuration
    are shown, including zero miss rates.}
\end{figure}

\myparagraph{Meeting both SLOs.}
At nominal load, time sharing misses a third of rendering deadlines;
redirect-only meets both SLOs in all five runs
(Figure~\ref{fig:vr}).
Under heavy load, their rendering miss rates reach 52.1\% and 78.1\%,
respectively, and every measured perception job is late.
With $q=20$, VUDA-full meets both SLOs: rendering misses are zero and
perception misses have a median of 2.78\% (2.64--3.75\% across runs).
Median per-run p99 latencies are 8.45 and 18.47\,ms, respectively.
No tested heavy-load setting meets a stricter 1\% limit for both tasks.

\begin{figure}[t]
  \centering
  \includegraphics[width=\columnwidth]{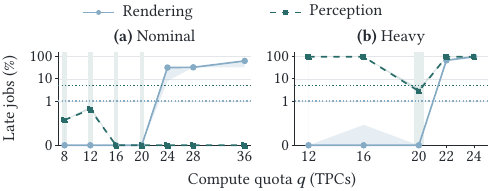}
  \caption{VR quota sensitivity. Points show medians and bands min--max
    over five runs; vertical shading marks tested quotas meeting both
    SLOs in every run. Dotted lines mark the limits.
    Vertical scaling follows Figure~\ref{fig:vr}; quota ranges differ.}
  \label{fig:vr-quota}
  \Description{Rendering and perception miss rates versus compute TPC
    quota. The successful tested quotas are 8, 12, 16, and 20 at nominal
    load, and only 20 at heavy load. Untested quotas are not marked as
    feasible.}
\end{figure}

\myparagraph{Choosing the partition.}
Both SLOs hold at tested quotas $q\in\{8,12,16,20\}$ under nominal
load, but only at $q=20$ under heavy load
(Figure~\ref{fig:vr-quota}).
At heavy load, $q=16$ protects rendering but misses every perception
deadline; $q=22$ violates both SLOs.
A separate four-point load sweep (two runs per point) also finds that
$q=20$ meets both limits across solo perception latencies of 4.6--8.1\,ms.
Partitioning enables joint SLO attainment, but higher loads require
more careful quota selection.

\section{Discussion \& Limitations}
\label{sec:discussion}

Our findings motivate more flexible work redistribution within the
graphics pipeline.
Beyond the redistribution between vertex and pixel processing described
in \S\ref{sec:hardware:anatomy}, additional redistribution points or
greater decoupling of pipeline stages from GPC-local resources could
allow work to follow available execution capacity.
This could enable finer-grained partitioning within the graphics
pipeline and relax the need for balanced per-GPC quotas
(\S\ref{sec:hardware:balance}).
Such flexibility would incur additional costs for redistributing work
and transferring pipeline state, presenting a trade-off for hardware
designers to explore.
The current design leaves L2 cache and device memory bandwidth shared
even with inter-SM partitioning, so \sys cannot eliminate bandwidth
contention between graphics and compute.
Our study is also limited to mainstream NVIDIA GPUs; we have not
investigated AMD or other GPU platforms.

\section{Related Work}
\label{sec:related}

\subsection{Sharing among Compute Workloads}

\myparagraph{Spatial partitioning.}
GPU sharing has primarily targeted compute tenants.
CUDA green contexts expose SM partitions through the driver
API~\cite{nvidia_green_contexts,nvidia_green_contexts_api}.
MPS~\cite{nvidia_mps} supports concurrent CUDA processes with resource limits,
while MIG provides hardware-isolated instances with predefined
profiles~\cite{nvidia_mig,nvidia_mig_profiles}.
Resource-provisioning systems build on these facilities to allocate capacity
among inference workloads~\cite{DBLP:conf/cloud/DhakalKR20,DBLP:journals/tpds/XuXCCSZL23,DBLP:conf/cloud/LiPSGT22}.
libsmctrl~\cite{bakita2023libsmctrl} exposes
launch-descriptor SM masks, and Bullet~\cite{DBLP:conf/asplos/0001XC0L026}
selects preconfigured SM-masked CUDA streams to spatially orchestrate prefill
and decode.
These mechanisms do not directly control native graphics placement.

\myparagraph{Kernel transformation and execution control.}
Tally~\cite{DBLP:conf/asplos/ZhaoJP25} combines slicing with persistent
thread blocks: worker blocks repeatedly fetch logical blocks and check
preemption signals between iterations.
MorphX~\cite{ren2026morphable} uses a micro-scheduler on each SM to dispatch
blocks, allowing running kernels to yield SMs and reclaim capacity
through additional kernel instances sharing the same work pool.
LithOS~\cite{10.1145/3731569.3764818} instead atomizes kernels to enable
fine-grained TPC reallocation.
REEF~\cite{280914} combines reset-based preemption of idempotent kernels with
dynamic kernel padding for low-latency sharing.
These approaches control compute kernels; native graphics shaders remain
coupled to the pipeline and fixed-function stages, rather than being exposed
as independently schedulable CUDA kernels.

\myparagraph{Software scheduling over time.}
Paella~\cite{DBLP:conf/sosp/NgD023} co-designs the compiler and runtime to
control kernel execution order.
Orion~\cite{strati2024orion} regulates kernel dispatch according to
interference, allowing compatible kernels to overlap.
XSched~\cite{shen2025xsched} exposes preemptible command queues across
accelerators, while GPREEMPT~\cite{fan2025gpreempt} combines timeslice-based
context switching with pre-preemption hints.
Higher-level systems multiplex GPU capacity across jobs and
containers~\cite{DBLP:conf/osdi/XiaoRLZHLFLJ20,DBLP:conf/osdi/XiaoBRSKHPPZZYZ18,258890,transparentgpusharing}.
These policies control ordering and responsiveness without spatially
partitioning native graphics and compute.

\subsection{Sharing between Compute and Graphics}

PilotFish~\cite{zhang2022pilotfish} harvests gaps between cloud-game frames,
and earlier driver- and hypervisor-level schedulers provide priorities,
reservations, or preemption for GPU commands
~\cite{capodieci2018deadline,DBLP:conf/usenix/KatoLRI11,DBLP:conf/usenix/KatoMMB12}.
Combo~\cite{chen2023combo} achieves spatial sharing on real GPUs using
transformed OptiX/CUDA rendering workloads.
KRYPTON~\cite{zhang2025krypton} uses time sharing in its evaluated Vulkan path.
CRISP~\cite{pan2024crisp} studies spatial graphics--compute interference
in simulation, without a deployable resource-control mechanism for real GPUs.
CUDA in Graphics (CiG)~\cite{nvidia_tensorrtrtx_cig} supports co-execution
on supported platforms and software versions.
NVIDIA's NVIGI interface exposes relative-priority
hints~\cite{nvidia_nvigi_gpusched}, and Vulkan interoperability is specified
separately~\cite{khronos_vk_nv_external_compute_queue}; neither interface
exposes an explicit, fine-grained graphics--compute resource split.
\sys joins native CUDA and Vulkan channels in one TSG with separate data
address spaces (\S\ref{sec:software}) and runtime-adjustable TPC partitions
(\S\ref{sec:hardware}), without rewriting kernels or shaders.

\section{Conclusion}
\label{sec:concl}

This paper demystifies the software and hardware barriers to
graphics--compute sharing on NVIDIA GPUs and presents \sys to overcome
them.
\sys combines channel redirection across CUDA and Vulkan with
runtime-adjustable TPC partitioning, preserving native execution stacks
without modifying drivers or rewriting kernels and shaders.
Evaluation across four application scenarios on three GPU platforms
shows that concurrency improves simulation throughput, while appropriate
partitioning enables SLO attainment where concurrency alone is
insufficient.

\bibliographystyle{ACM-Reference-Format}
\bibliography{references}

\appendix

\end{document}